\documentclass{PoS}

\usepackage{units}
\usepackage{url}
\usepackage{upgreek}

\title{Simulation of the expected performance for the proposed gamma-ray detector HiSCORE}

\ShortTitle{Simulation of the expected performance for the proposed gamma-ray detector HiSCORE}

\author{Daniel Hampf, Martin Tluczykont, Dieter Horns%
        \\
       Department of Physics, University of Hamburg, Luruper Chaussee 149, 22761 Hamburg, Germany\\
       E-mail: \email{daniel.hampf@desy.de}}

\abstract{The HiSCORE project aims at opening up a new energy window in gamma-ray astronomy: The energy range above 30 TeV and up to several PeV. For this, a new detector system is being designed. It consists of a large array of non-imaging Cherenkov detectors with a light sensitive area of 0.5 square metres each. The total effective area of the detector will be 100 square kilometres. A large inter-station distance of 150 metres and a simple and inexpensive station design will make the instrumentation of such a large area feasible.

A detailed detector simulation and event reconstruction system has been developed and used in conjunction with the CORSIKA air-shower simulation to estimate the sensitivity of the detector to gamma-ray point sources. The threshold for gamma-rays is 44 TeV (50\% trigger efficiency) in the standard configuration, and the minimal detectable flux from a point source is below $\unit[10^{-13}]{erg\ s^{-1} \ cm^{-2}}$ above 100 TeV.

Several options to lower the energy threshold of the detector have been examined. The threshold is decreased to 34 TeV by a smaller station spacing of 100 metres, and is further decreased to 24~TeV if the detector is set up at an altitude of 2000 metres above sea level. At a spacing of 150 metres, however, a higher altitude has no positive effect on the energy threshold.

The threshold can also be reduced if the detector stations consist of small, independent 2 x 2 sub-arrays. In this case, the station spacing is not significant for the threshold, but again a higher altitude decreases the threshold further.

}

\FullConference{25th Texas Symposium on Relativistic Astrophysics \\
		 December 6-10, 2010\\
		 Heidelberg, Germany}

\begin{document}

\section{Introduction}
Ground-based gamma-ray astronomy has become a thriving discipline and a wealth of new discoveries has been made in recent years  - see e.g.\ \cite{2008RPPh...71i6901A} or \cite{2008RvMA...20..167H} for reviews. The key instruments for this research are Cherenkov Telescopes like MAGIC \cite{2008ApJ...674.1037A}, HESS \cite{2006A&A...457..899A}, and VERITAS \cite{2008ApJ...679.1427A}. 
Only the large effective areas of up to $\unit[10^5]{m^2}$ achieved by Cherenkov Telescopes make the detection of weak gamma-ray fluxes in the very high energy band (VHE, $\unit[30]{GeV}$ to $\unit[30]{TeV}$) possible. 

However, beyond about $\unit[30]{TeV}$ (ultra high energy band, UHE) the flux of almost all known gamma-ray sources is too weak to be detected by current instruments, and even larger effective areas are needed.  
The HiSCORE detector is specifically designed to fill this gap with an effective area of $\unit[100]{km^2}$. Using small non-imaging Cherenkov detectors with a large inter-station spacing, it becomes feasible to equip such a large area at a reasonable effort. A detailed description of the detector will be given in section \ref{detector}.

Many interesting open questions in high energy astronomy can be addressed by observations in the UHE band, above all the mystery of the acceleration of charged cosmic rays. Gamma-ray observations in the UHE band can help to unambiguously identify sites of hadronic cosmic ray acceleration, since there is a significant drop in the efficiency of leptonic processes (inverse Compton effect) at these energies. This and other scientific objectives of gamma-ray astronomy in the UHE band are outlined in \cite{Tluczykont_COSPAR}.

In order to estimate the instrument's sensitivity to UHE gamma-rays, a detailed detector simulation has been developed, including an event reconstruction algorithm. This has been used to derive key performance figures of the detector like angular resolution, energy resolution and particle separation power, and to calculate the sensitivity to gamma-ray point sources in the UHE regime. This simulation and its results will be presented in section \ref{simulation}.

The HiSCORE project aims at achieving an overlap in energy with existing and other planned gamma-ray observatories for cross-checks and to obtain continuous spectra of gamma-ray sources. However, a threshold in the low TeV regime is challenging for a non-imaging detector like HiSCORE. Section \ref{threshold} presents possible modifications of the standard detector configuration that may help to lower the energy threshold.

\section{The HiSCORE detector}
\label{detector}

The HiSCORE detector will consist of a regular grid of non-imaging wide-angle Cherenkov detectors with an inter-station spacing of about $\unit[150]{m}$. A similar approach has been pursued by the HEGRA collaboration with the AIROBICC detector, however with a significantly smaller total effective area of only $\unit[3 \times 10^4]{m^2}$ \cite{1995APh.....3..321K}.

\begin{figure}[tb]
	\begin{minipage}[b]{0.3\linewidth}
		\centering
		\includegraphics[width=\textwidth]{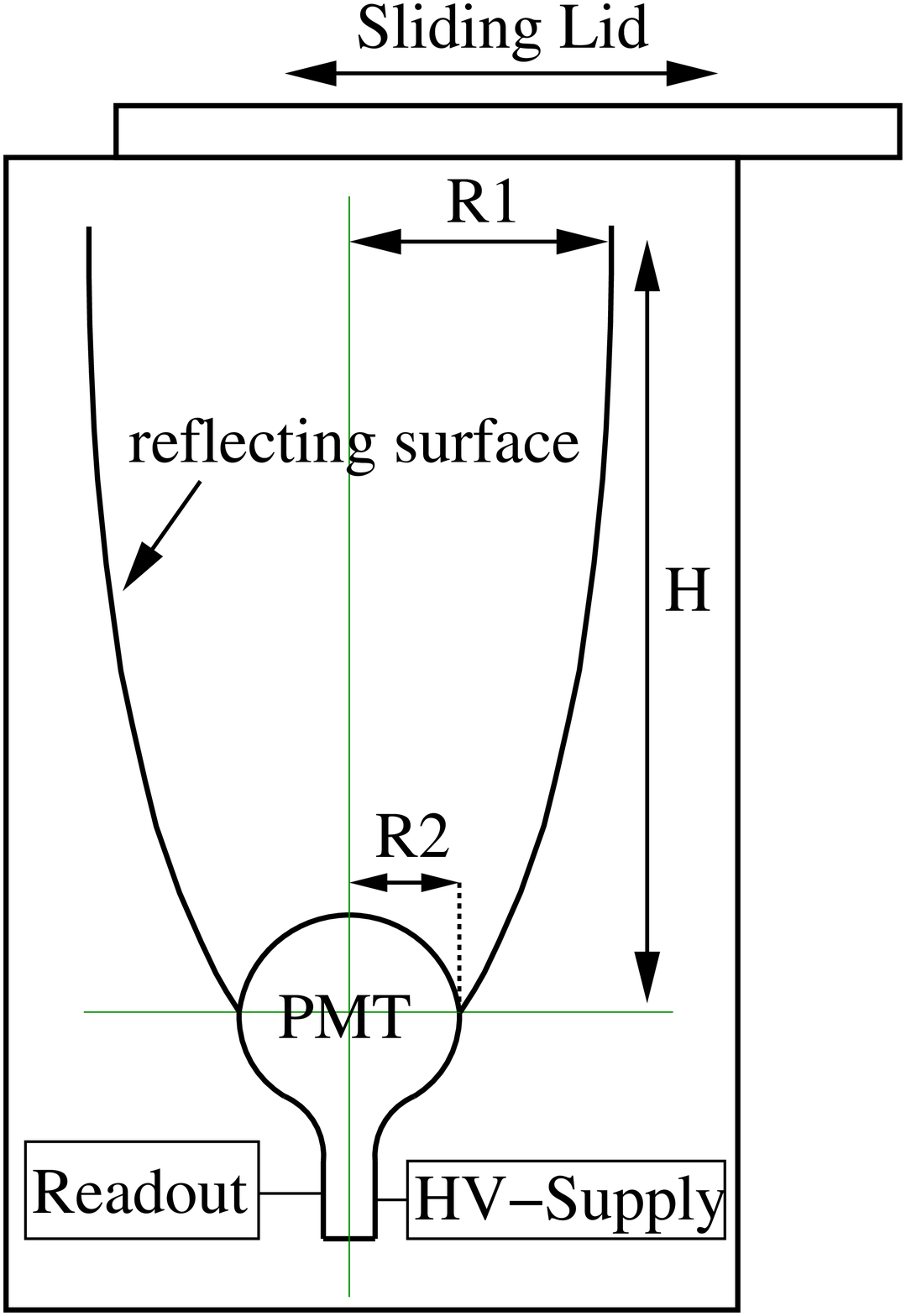} 
		\caption{Layout of one detector module.  R$1 = \unit[10]{cm}$,  \mbox{R$2 = \unit[20]{cm}$}, H$ = \unit[52]{cm}$}
		\label{station}
	\end{minipage}
	\hspace{0.5cm}
	\begin{minipage}[b]{0.65\linewidth}
		\centering
		\includegraphics[width=\textwidth]{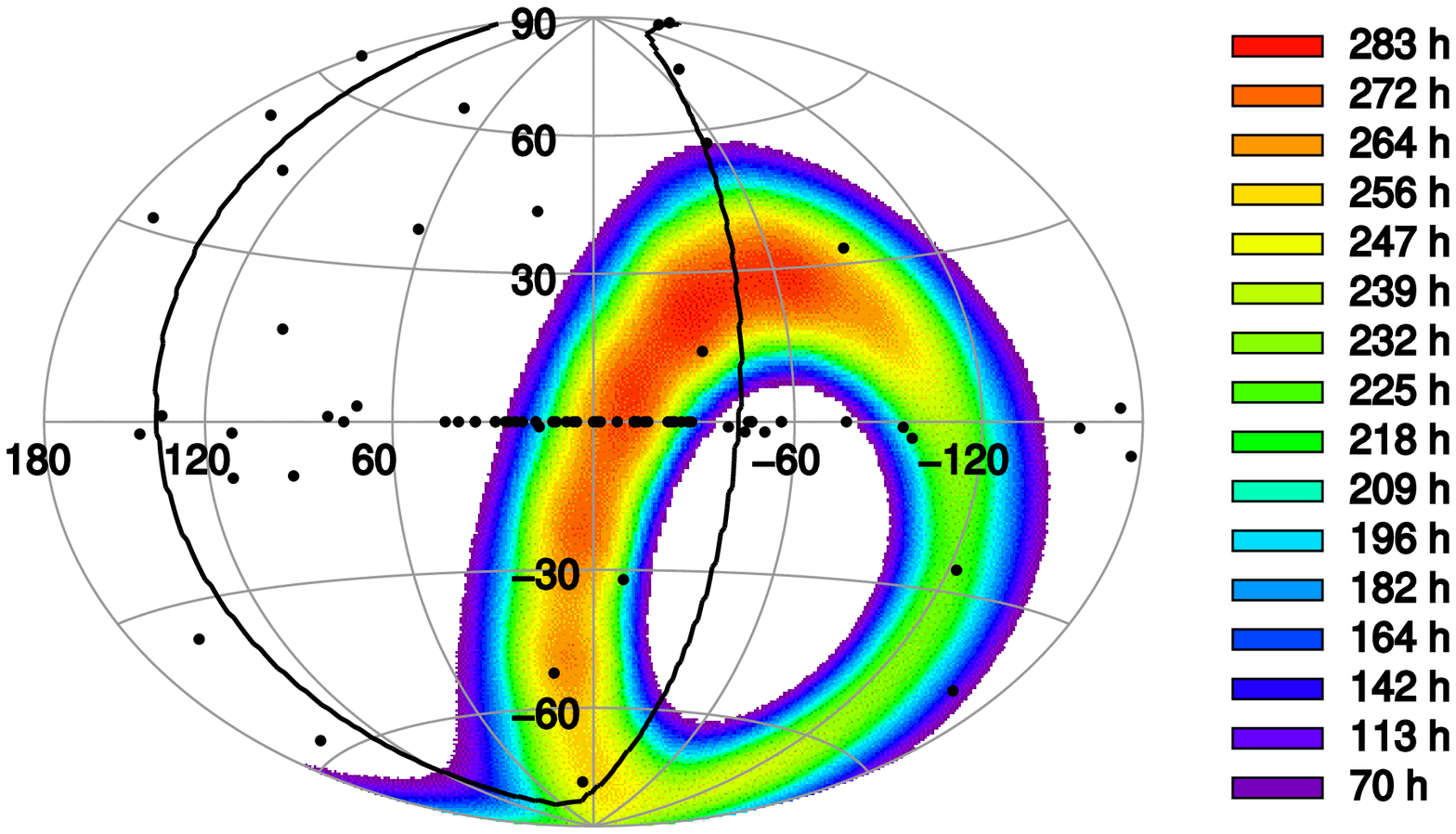} 
		\caption{Visible regions of the sky for HiSCORE from a southern hemisphere site (35$^\circ$ south). Given are hours of exposure per year, without bad weather correction, in galactic coordinates. Small black dots are known VHE gamma-ray sources, the black line shows the position of the super-galactic plane.}
		\label{skymap}
	\end{minipage}
\end{figure}

Each detector station contains four detector modules, each of which consists of an 8'' photomultiplier with a light concentrator on top, facing towards zenith (see figure \ref{station}). The light concentrator (Winston cone) has a half opening angle of $30^\circ$ which allows to detect air-showers up to at least $25^\circ$, resulting in a simultaneous monitoring of about 0.6 sr of the sky. The light-sensitive area of a four-module station is $\unit[0.5]{m^2}$.

The signals of the four channels are summed up and stored by a fast digital readout system like the DRS4 chip\footnote{\url{http://drs.web.psi.ch/}}. The trigger is constructed locally using the clipped sum signal of the four modules in a similar fashion as the sum trigger in MAGIC \cite{2008Sci...322.1221A}. On the first level, each station operates independently in order to avoid the need for fast inter-station communication. All signals above threshold are saved and sent to a central computer system over a standard wireless network connection. An event is accepted if three stations have triggered and the shower core position is reconstructed to be within the boundaries of the detector array (``acceptance cuts'').

In order to reconstruct air-shower events, each data package that is sent to the central computer needs to contain an accurate time-stamp. A system using GPS and radio wave interferences is currently under development. In a first stage, the detector electronics and communication will be controlled by a small PC at each station; however, in the long term most of the electronics will be incorporated into an FPGA.

It is currently planned to deploy the detector on a southern hemisphere site in order to allow direct observation of a large part of the inner Galactic region. Figure \ref{skymap} shows the expected one-year sky exposure times for a site at $35^\circ$ south (southern Australia, Namibia, Argentina, etc.). About 35\% of the sky are in the field of view for more than 100 hours per year, and 26\% (including the central Galaxy region) are viewed for more than 200 hours per year. 


\section{Simulation of detector performance}
\label{simulation}
A detailed detector simulation has been developed and tested using simulated air-showers from CORSIKA with the IACT option \cite{heck:1998a:long}. The detector simulation contains atmospheric absorption, angle- and wavelength dependent transmission of the light concentrator, photomultiplier quantum efficiency, pulse shaping and all relevant noise sources, especially noise generated by night sky brightness. 

The output of the detector simulation is analysed by an event reconstruction algorithm that reconstructs shower core position, direction and energy of primary particle and the position of the shower maximum. The event data is also used to estimate the nature of the primary particle, i.e. to distinguish photons from hadrons and to study the mass composition of the cosmic rays.

\begin{figure}[tb]
	\begin{minipage}[b]{0.49\linewidth}
		\centering
		\includegraphics[width=\textwidth]{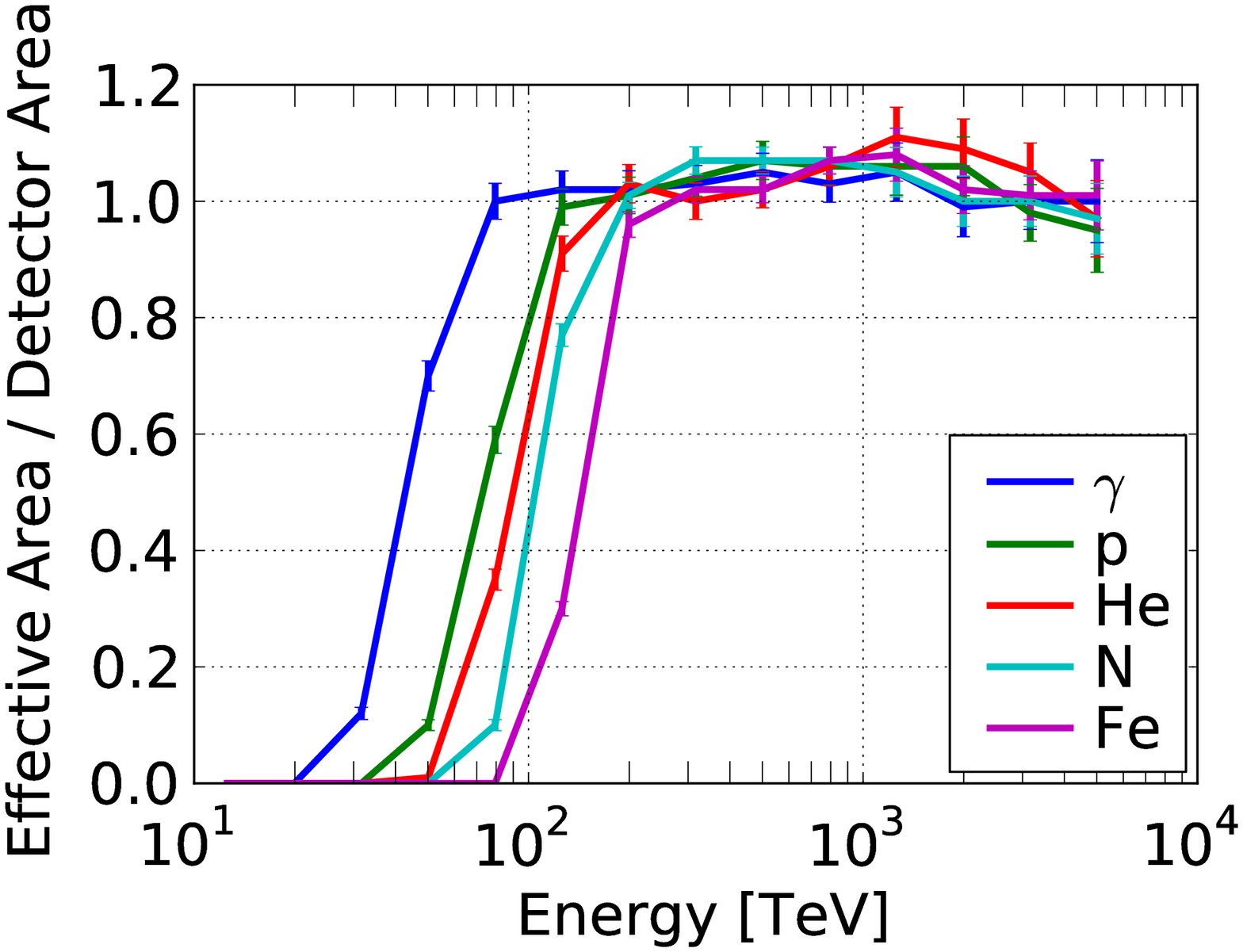} 
		\caption{Effective areas for gamma-rays and four types of hadronic cosmic rays after acceptance cuts.}
		\label{eff_areas_cutB}
	\end{minipage}
	\hspace{0.02\textwidth}
	\begin{minipage}[b]{0.49\linewidth}
		\centering
		\includegraphics[width=\textwidth]{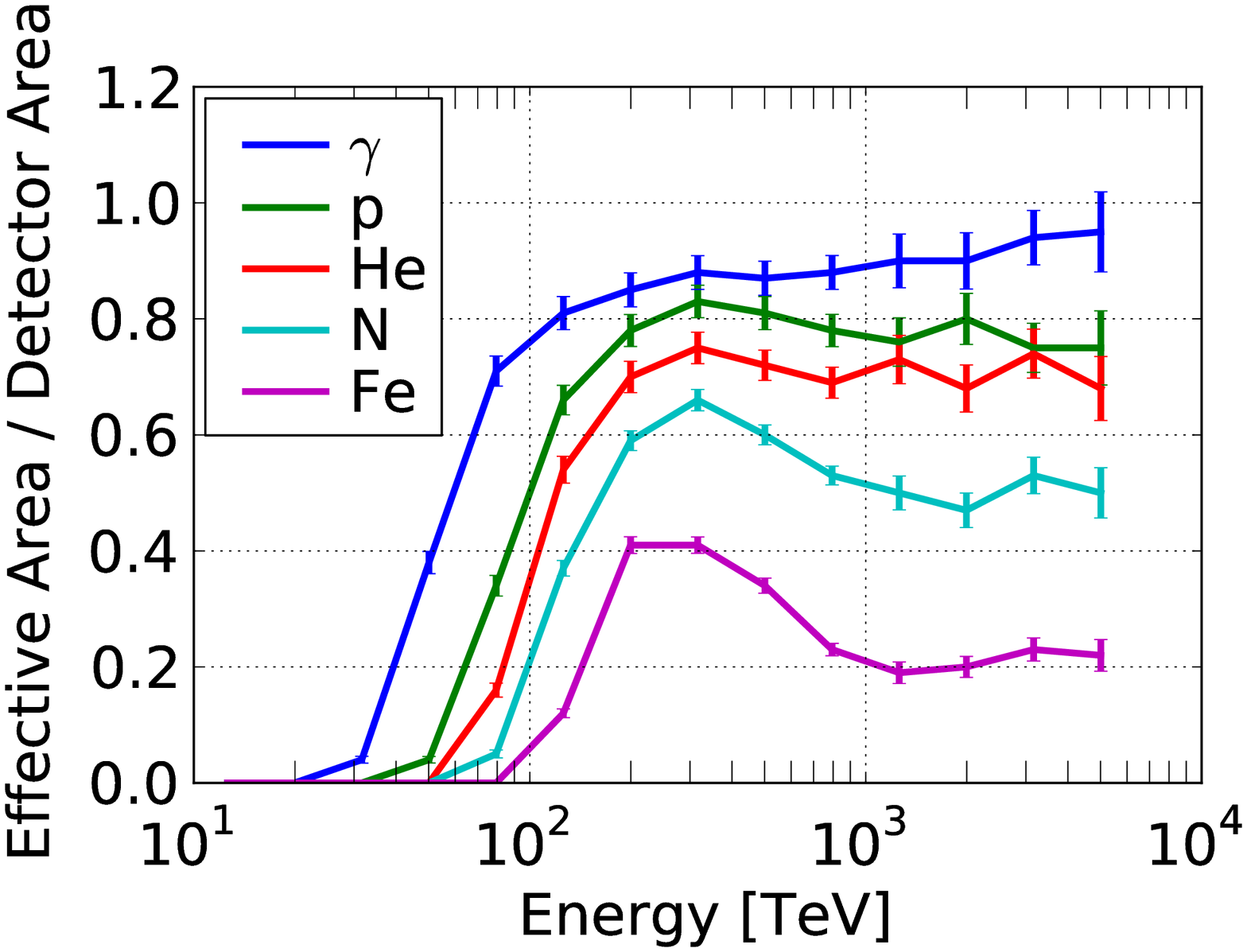} 
		\caption{Effective areas for gamma-rays and four types of hadronic cosmic rays after gamma-ray cuts.}
		\label{eff_areas_cutC}
	\end{minipage}
\end{figure}

The resolution of the reconstructed values depends strongly on the number of triggered stations, and therefore on the energy of the primary particle. The angular resolution is about 0.8$^\circ$ (68\% containment) at the threshold and 0.1$^\circ$ at $\unit[200]{TeV}$, while the energy resolution is 40\% and 10\% respectively.
The energy threshold for gamma-rays is $\unit[44]{TeV}$ (50\% efficiency) and about $\unit[150]{TeV}$ for iron nuclei. The effective areas after acceptance cuts are shown in figure \ref{eff_areas_cutB}. 

\begin{figure}[tb]
	\centering
	\includegraphics[width=0.8\textwidth]{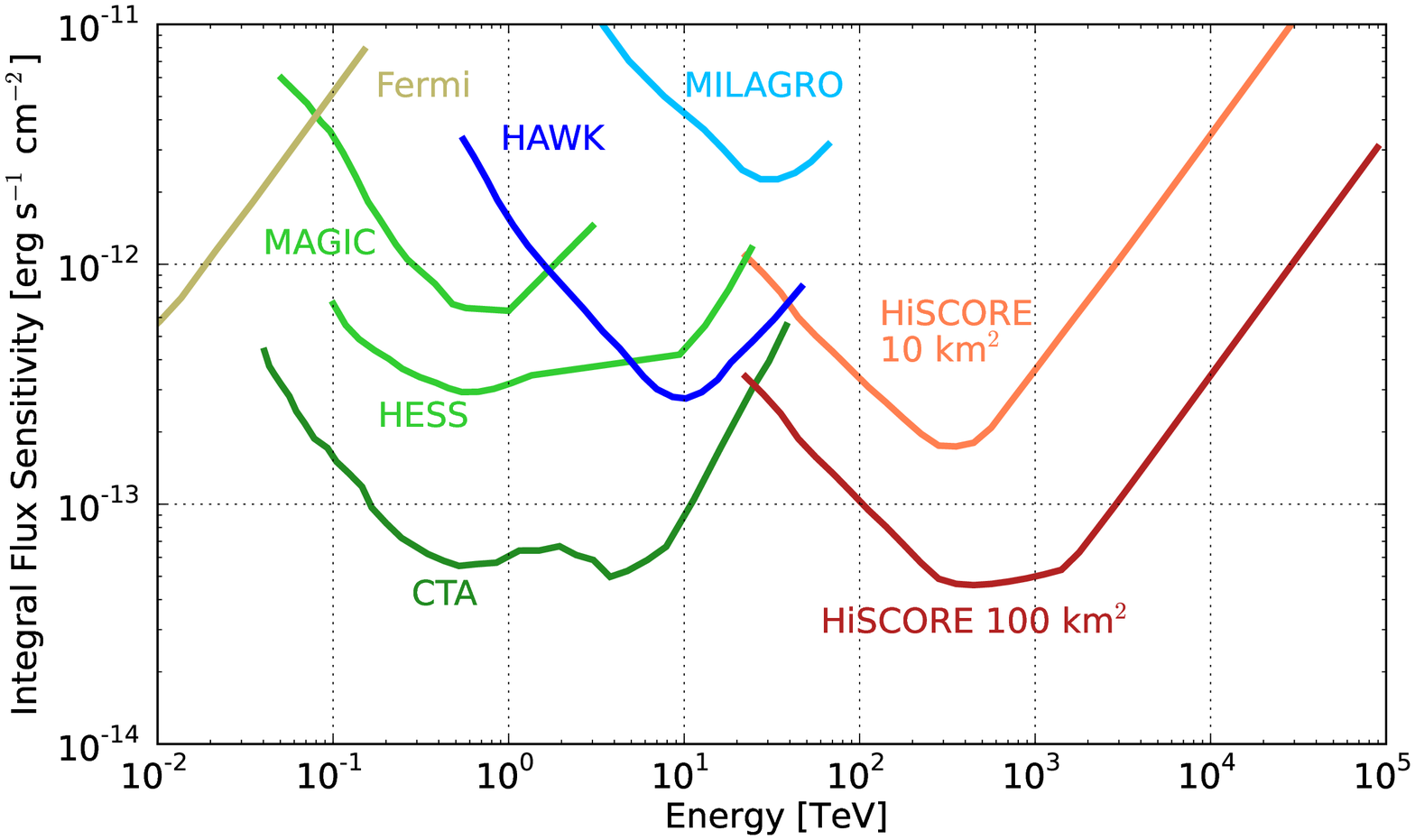} 
	\caption{Integral flux sensitivity of selected current and future gamma-ray observatories. For HESS, MAGIC and CTA, 50 hours of pointed observation is assumed. For all other instruments, continuous 5 year operation in survey mode is assumed, which corresponds to about 1000 hours in the case of HiSCORE. References for instrument sensitivities are: \cite{2006A&A...457..899A}, \cite{2005ICRC....5..227B}, \cite{2009ApJ...697.1071A}, \cite{2009ICRC_bernloehr}, \cite{2008RPPh...71i6901A}}
	\label{sensitivities}

\end{figure}

After the application of a gamma / hadron separation algorithm, heavy hadrons are efficiently suppressed (see figure \ref{eff_areas_cutC}). The angular resolution, the effective areas after the gamma / hadron separation and the cosmic ray rates from \cite{hoerandel:2003a} are used to estimate the sensitivity of the HiSCORE detector\footnote{The sensitivity is defined as the minimal flux at which a gamma-ray point source can be detected above the background of cosmic rays with a significance of $5\upsigma$, with a minimum of 50 gamma events as additional requirement.}. Figure \ref{sensitivities} shows the sensitivity of a $\unit[10]{km^2}$ and a $\unit[100]{km^2}$ HiSCORE detector array in comparison to other existing and planned gamma-ray observatories. 


\section{Concepts to lower the energy threshold}
\label{threshold}
In this section, several options to decrease the threshold to below the $\unit[44]{TeV}$ achieved in the standard configuration have been examined: A change of the observation altitude, a closer spacing of the detector stations and the use of small 2 x 2 sub-arrays instead of standard detector stations.

At higher altitudes the Cherenkov light is more concentrated around the shower core position than at lower altitudes, which results in a higher signal in the central detector stations and smaller signals further away from the core position.
Simulations were carried out for a detector at sea level and a detector at \unit[2000]{m} a.s.l.\ to examine the consequences of this effect on the energy threshold.

\begin{figure}[tb]
	\begin{minipage}[b]{0.49\textwidth}
		\centering
		\includegraphics[width=\textwidth]{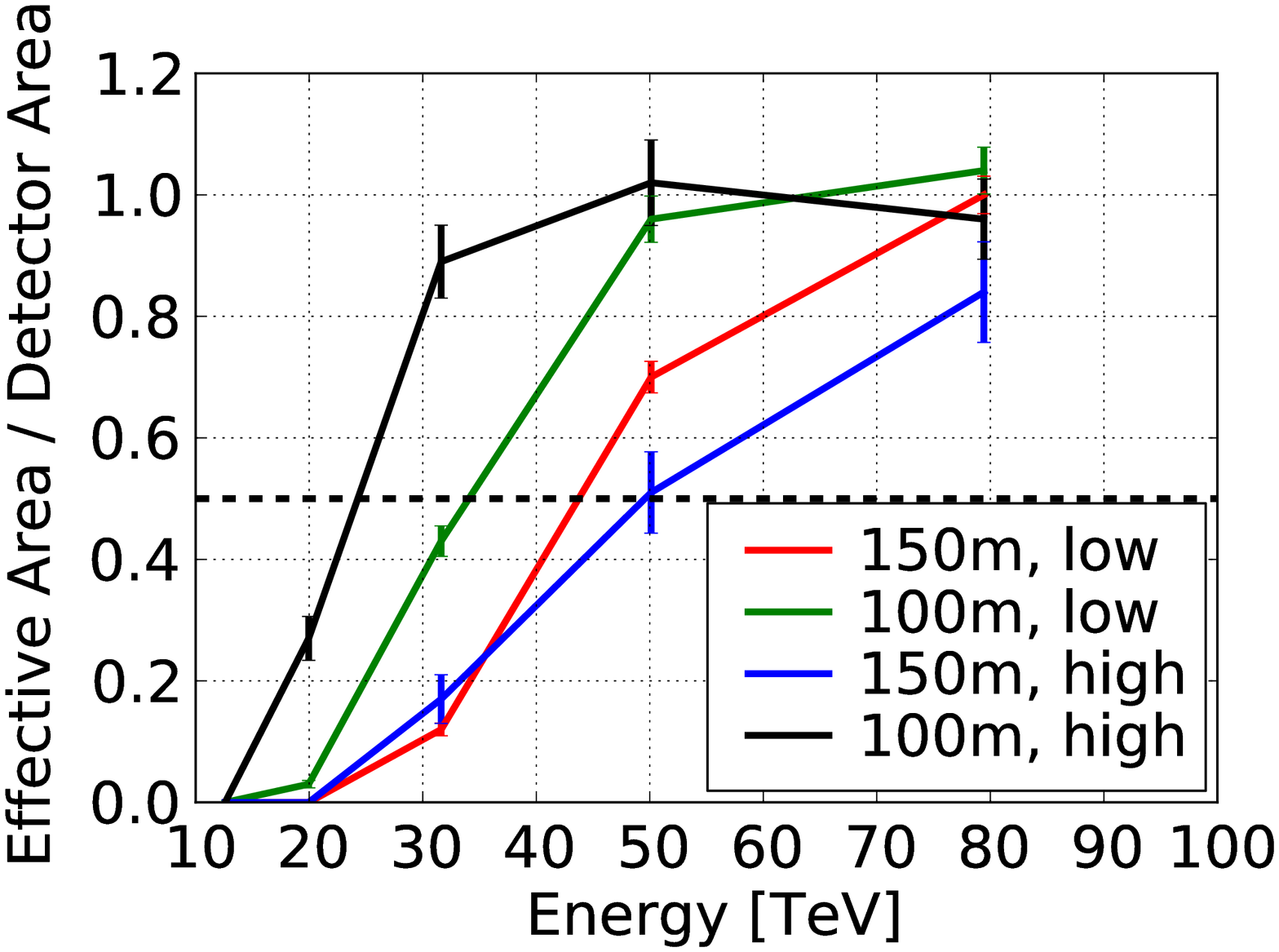} 
  	    \caption{Effective areas for a detector array at sea level ("low") and at \unit[2000]{m} \mbox{a.s.l.} ("high") with an inter-station spacing of \unit[100]{m} and \unit[150]{m}.}
		\label{comparison_cutB}
	\end{minipage}
	\hspace{0.02\textwidth}
	\begin{minipage}[b]{0.49\textwidth}
		\centering
		\includegraphics[width=\textwidth]{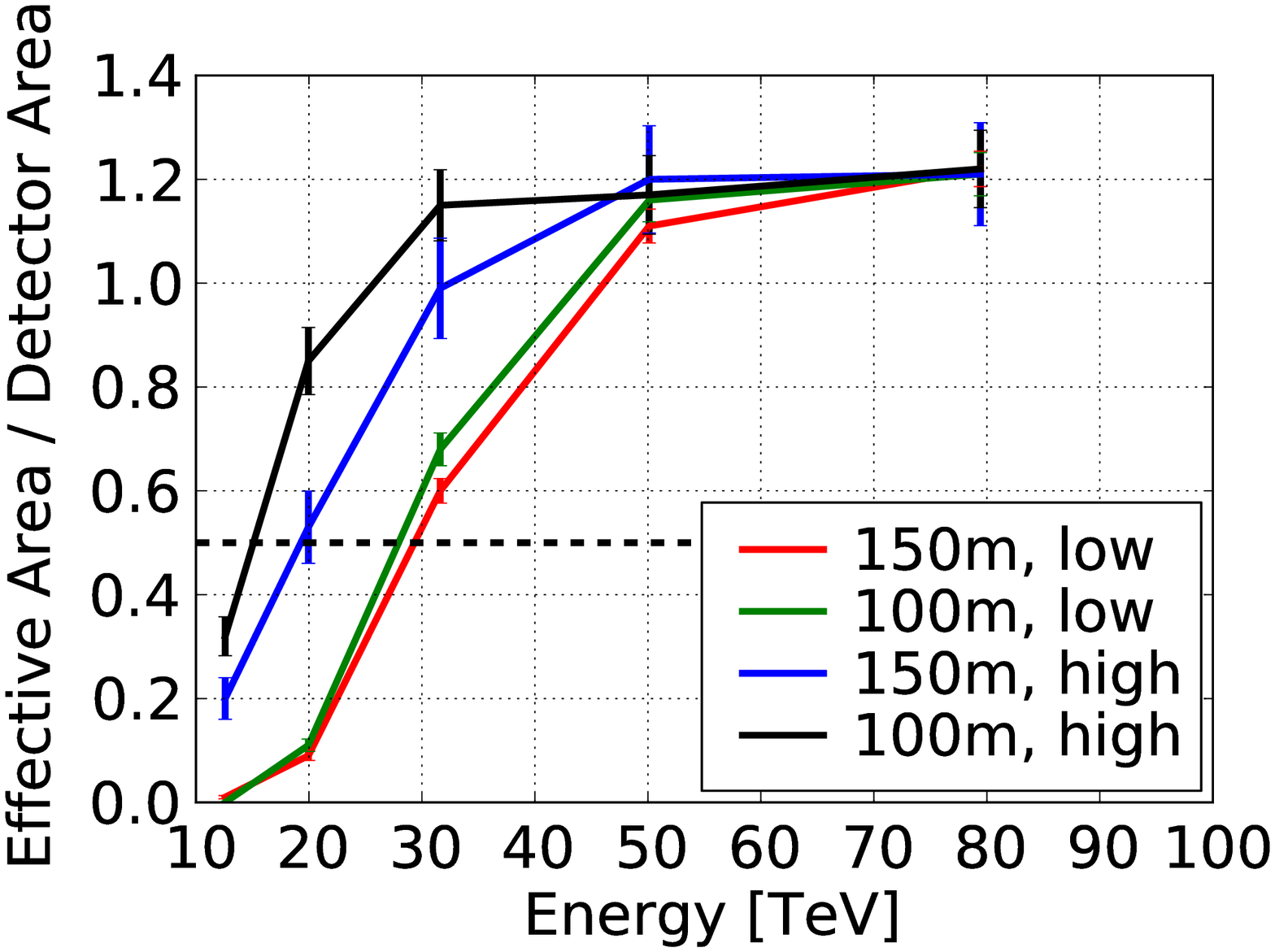} 
		\caption{Effective areas for a detector array with \mbox{2 x 2} sub-arrays at sea level and at \unit[2000]{m} \mbox{a.s.l.} with a station spacing of \unit[100]{m} and \unit[150]{m}.}
		\label{comparison_T1}
	\end{minipage}
\end{figure}

Figure \ref{comparison_cutB} shows the effective areas for gamma-rays after acceptance cuts for a detector array with a spacing of \unit[100]{m} and one with a spacing of \unit[150]{m}, both at high and low altitude. For a spacing of \unit[150]{m} (the standard configuration), a higher altitude has no positive effect on the threshold, which is about \unit[44]{TeV} (50\% trigger efficiency) at sea level and \unit[50]{TeV} at \unit[2000]{m} a.s.l. For a detector with  \unit[100]{m} spacing, the threshold is lowered from \unit[34]{TeV} to \unit[24]{TeV} by going to a higher altitude. Generally, the threshold is lower for the detector with a closer spacing, as on average more stations are triggered at a given energy.

A second strategy to lower the threshold is to split up the four modules of a standard detector station into four separate detectors about five to ten metres apart from each other. These 2 x 2 sub-arrays can perform a basic reconstruction independently from each other, so that a single triggered sub-array is enough to accept an event. The trigger threshold is generally lower in this configuration (see effective areas in figure \ref{comparison_T1}), and the spacing of the sub-arrays has not a strong influence on the threshold (because near the threshold most events are detected only by a single sub-array). At low altitude the threshold is about \unit[28]{TeV} for both spacings, while the threshold at \unit[2000]{m} a.s.l. is \unit[19]{TeV} for the \unit[150]{m} spacing and \unit[15]{TeV} for the \unit[100]{m} spacing. In this scenario, the higher altitude decreases the threshold significantly, since the steeper lateral distribution function of the Cherenkov light has no negative effect on the single-sub-array events. It has to be noted, however, that single-sub-array events cannot be reconstructed as accurately as usual accepted events, and the sensitivity in the single-sub-array regime may be considerably worse than in the three stations regime.

\newpage

\section{Conclusions and Summary}
The presented simulations show that the HiSCORE detector concept is well capable of opening up the observation window of the ultra-high energy band of gamma-ray astronomy. Point sources can be detected down to a flux level of $\unit[10^{-13}]{TeV}$ by a $\unit[100]{km^2}$ array. The simple design of the individual stations and the large inter-station spacing allows the instrumentation of such a large area at a reasonable effort.

During its nominal operation time of five years, the detector continuously scans a large fraction of the sky, including promising regions for ultra-high energy gamma-ray sources like the Galactic Centre region. About 35\% of the sky are in the field of view for more than 100 hours per year.

The reconstruction techniques developed so far allow for a direction reconstruction with an accuracy of better than 0.1$^\circ$, an energy reconstruction of better than 10\% and a gamma / hadron separation that is also capable of separating different charged cosmic rays for composition studies.

The gamma-ray threshold of a detector in standard configuration (\unit[150]{m} spacing, sea level) is about \unit[44]{TeV} (50\% efficiency), which results in a small overlap with some existing and planned detector systems. A closer inter-station spacing of \unit[100]{m} or the use of \mbox{2 x 2} sub-arrays are promising concepts for lowering the threshold. In both scenarios a higher detector altitude helps to further decrease the detector threshold, while there is no benefit of a higher altitude when using the standard layout.

Both concepts, a closer spacing and a detector layout with sub-arrays, result in a higher cost of the detector system per area, so that a detector with a lower threshold will cover a smaller total area. A trade-off between a low energy threshold and a good overall sensitivity has to be made in order to decide on the final detector configuration.


\section*{Acknowledgements}
Daniel Hampf likes to thank the German Ministry for Education and Research (BMBF) for its financial support (contract number 05A08GU1).

\bibliographystyle{unsrt}
\bibliography{TEXAS2010_proceedings}

\end{document}